# Concept-oriented model:
# inference in hierarchical multidimensional space

Alexandr Savinov

Database Technology Group, Technische Universität Dresden, Germany

http://conceptoriented.org

**Abstract**

In spite of its fundamental importance, inference has not been an inherent function of multidimensional models and analytical applications. These models are mainly aimed at numeric (quantitative) analysis where the notions of inference and semantics are not well defined. In this paper we argue that inference can be and should be integral part of multidimensional data models and analytical applications. It is demonstrated how inference can be defined using only multidimensional terms like axes and coordinates as opposed to using logic-based approaches. We propose a novel approach to inference in multidimensional space based on the concept-oriented model of data and introduce elementary operations which are then used to define constraint propagation and inference procedures. We describe a query language with inference operator and demonstrate its usefulness in solving complex analytical tasks.

## 1. Introduction

Euclidean geometry is an axiomatic system deriving various geometric properties as *theorems* from a small number of *axioms*. It dominated for more than 2000 years until René Descartes revolutionized mathematics by developing Cartesian geometry (also known as analytic geometry) by introducing the formal notions of *variable* and *coordinates*. The astounding success of analytical geometry was due to its ability to reason about geometric objects and analyze their properties numerically which turned out to be more practical and intuitive in comparison with formal logic.

The area of data modeling and analysis can also be characterized as having two branches or patterns of thought. The first one follows the Euclidean axiomatic approach where data is described using propositions, predicates, axioms, inference rules and other formal logic constructs. For example, deductive data models and the relational model are based on the first-order logic where a database is represented as a number of predicates. The second branch in data modeling relies on the Cartesian conception where data is thought of as a number of points in a multidimensional space with properties represented as coordinates along its axes. This multidimensional approach has been proven to be extremely successful in analytical applications, data warehousing and OLAP.

One negative consequence of having two different views on data is that a typical data management system consists of two components: one for *transactional* operations and the other for *analytical* operations. In addition, many systems have also a semantic and reasoning component. These two subsystems have different designs, serve different purposes, and use different methods but rely on the same data and constitute one system. Essentially, one and the same data exists simultaneously in several worlds and the need to maintain them results in numerous problems at all stages of the enterprise data management system life-cycle.

Multidimensional models have been around for a long time and a variety of approaches have been proposed for representing multidimensional data (Agrawal et al., 1997; Gyssens & Lakshmanan, 1997; Li & Wang, 1996; Nguyen et al., 2000; Pedersen & Jensen, 2001; Pedersen, 2009). Yet, all of them have one serious drawback: it is not clear how to represent data semantics and how to reason about data in it. These are essential functions of a database because they allow it "to respond to queries and other transactions in a more intelligent manner" (Codd, 1979). A database with reasoning capabilities could automatically derive relevant data in one part of the model given constraints in another part without the need to specify *how* it has to be done. For example, if we need to retrieve a set of writers by specifying only a set of publishers then this could be represented by the following query:

```
GIVEN (Publishers WHERE name == 'XYZ')
GET (Writers)
```

Importantly, this query does not have any indication of what is the schema and how it has to be processed. The database has to be smart enough to infer the requested data taking into account the imposed constraints and the data semantics.

Answering such queries is especially important for self-service analytics, agile analytics, analysis of open data with complex schemas (Thiele & Lehner, 2012). It is a highly non-trivial task and currently existing solutions rely on data semantics (Peckham & Maryanski, 1988), inference rules in deductive models (Ullman & Zaniolo, 1990) or structural assumptions as it is done in the universal relation model (URM) (Vardi, 1988). Yet, to the best of our knowledge, no concrete attempts to exploit multidimensional space for inference have been reported in the literature, apart from our preliminary results described in (Savinov, 2006).

This paper is a full version of the short paper (Savinov, 2012b). We present a novel solution to the problem of inference which relies on the multidimensional structure of the database. This approach allows us to not only do complex numeric analysis but also perform tasks which have always been a prerogative of logic-based models. The proposed approach is very simple and natural in comparison to logic-based approaches because it relies on only what is already in the database: dimensions and data.

The solution is based on a new general-purpose model, called the concept-oriented model (COM) (Savinov, 2014b, 2014c, 2012c, 2011a, 2009a). It provides a unified view on data by combining many existing views and patterns of thoughts currently used in data modeling. It is also tightly integrated with a novel approach to programming, called concept-oriented programming (COP) (Savinov, 2005c, 2008, 2009b, 2012a). A distinguishing property of this model is that it relies on order-theoretic basis (Savinov, 2014c) but its use of partial order is different from the approaches described in (Raymond, 1996) and (Buneman et al., 1991; Zaniolo, 1984). COM is inherently multidimensional and analytical but at the same time it works directly with transactional data without the need to define such elements as cubes, measures and dimensions for each analysis scenario like it is done in standard OLAP models (Pedersen & Jensen, 2001; Pedersen, 2009). It provides two operations of projection and de-projection which are then used to define constraint propagation and inference procedures.

More specifically, the paper makes the following contributions:

- We introduce a novel formal setting for describing multidimensional spaces which is based on *nested partially ordered sets*

- We define operations of *projection* and *de-projections* which serve as a basis for inference
- We define procedures of *constraint propagation* and *inference* as well as describe how they are supported by the query language
- We describe how this approach has been implemented as a translator to an in-memory engine for complex analytical computations

The paper has the following layout. Section 2 provides a background and describes related work. Section 3 illustrates the general idea of the proposed approach by considering a 2-dimensional space. Section 4 describes the formal setting by defining the notion of nested partially ordered set and how it is used to represent hierarchical multidimensional spaces. Section 5 defines main operations on nested posets and how they are used for inference. Section 6 makes concluding remarks.

## 2. Background and Related Work

Inference can be implemented in two major ways:

- Logic-based approaches
- Structure-based approaches

Logic-based approaches use one or another variant of formal logic like the first-order logic which is probably the most wide spread logic-based formalism in computer science. In particular, this approach is used in deductive databases (Ullman & Zaniolo, 1990), semantic data models (Peckham & Maryanski, 1988) and description logics which is a formal basis for ontologies and the Semantic Web. In the context of this paper, one of the most important properties of the logic-based approaches is the use of inference rules. They are used by any inference engine and determine what will be derived and how it will be derived. Inference is not possible without defining inference rules and different inference rules will result in different results.

For example, assume that there are a number of data sets represented as RDF triples. If we now select some elements and then try to propagate this constraint to another part of the model then it cannot be done automatically just because there are many ways how it can be performed. In order to carry out inference the engine needs inference rules without which constraint propagation process is ambiguous (except for some simple cases). Indeed, an RDF store can be represented as a graph and inference on a graph is an ambiguous procedure just because the roles of its elements (vertexes and edges) are not clearly (semantically) defined. This can be demonstrated by a simple graph shown in Fig. 1. If we select elements *b* and *c* as source constraint then it is not clear what can be said about other elements in this graph. In particular, will elements *e* and *f* be included in the result set? Inference rules are precisely what make inference possible in this case.

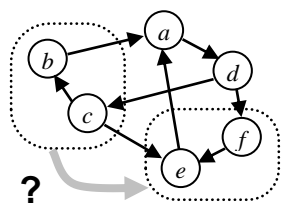

*Figure 1. Inference on a graph*

The second major approach employs structural aspects of the database to derive new data using imposed constraints. This approach is used in the universal relation model (URM) (Kent, 1981; Fagin et al., 1982) which "aims at achieving complete access-path independence in relational databases by relieving the user of the need for logical navigation among relations" (Maier, 1984). This goal is achieved by regarding the database as a single relation, called the universal relation, so that all other relations are its projections. Queries are written in terms of only attributes and the system uses the universal relation for reasoning about dependencies and translating the queries into the logical structure. For example, in System/U (Vardi, 1988) all banks where Jones is a customer are retrieved using its Quel language without specifying how it has to be done:

```
retrieve (BANK)
where CUST = "Jones"
```

A similar query for retrieving all order characterized by some properties can be written in the FIDL language (flexible interrogation and declaration language):

```
LIST ALL ORDERS
FOR WHICH PART IS SCREW OR BOLT
```

Importantly, the constrained properties are not directly connected with the elements we want to retrieve and the database engine relies on the structural rules of the URM to translate them into more specific queries.

The assumption of universal relation was shown to be incompatible with many aspects of the relational model and did not result in a new foundation for data modeling. Yet, the problem of abstracting from the database structure addressed by URM is still highly actual especially in the context of semantic databases and reasoning. COM is similar to URM because both rely on data structure for solving this problem. The main difference of COM is that it employs the formalism of partially orders sets for that purpose. URM has only two levels in its structure: attributes and relations (Fig. 2). The task is to propagate constraints between attributes using the available relations. One problem is that there are many alternative ways how to navigate from source to target attributes using many intermediate joins. For example, if there are three relations `Suppliers`, `Parts` and `SP` with the structure shown in Fig. 2 then there are at least two ways for propagating constraints from `Suppliers` to `Parts`: (i) by using the intermediate `SP` relation and two joins, and (ii) by joining these two relations on their `CITY` attribute. In more complex schemas, the number of possible propagation paths is much higher and the choice of a natural path and their interpretation becomes very difficult.

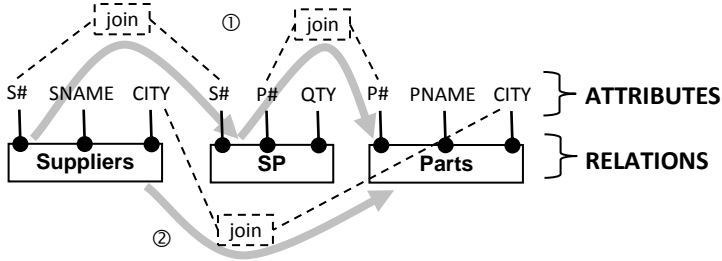

*Figure 2. Inference in URM*

Traditional multidimensional models also belong to structural methods of data analysis. In these models, all data elements are categorized as belonging to dimensions, facts or measures which can be viewed as roles data is playing during analysis. Yet, this approach has been mostly applied to numerical analysis like grouping and aggregation as opposed to logical analysis like inference. In the context of this paper, COM can be viewed as a generalization of multidimensional models by extending them with logical operations. This goal is achieved by removing the roles of dimensions, facts and measures from the model definition and treating them as relative roles defined for each specific analysis scenario. Grouping of facts around cube cells is performed similar to the traditional OLAP analysis. However, instead of applying a numerical aggregation function to grouped elements, COM skips this step and finds a projection of the group onto some target. In OLAP scenarios, the target is some numeric domain while in logical tasks it is normally a discrete domain similar to dimensions. For example, if a cell of a cube has an associated group of books then a typical OLAP analysis could find their average price while constrain propagation task could find all publishers of these books.

An interesting approach to combining quantitative and qualitative methods as it was proposed in (Dau & Sertkaya, 2011). This approach is based on using formal concept analysis (FCA) for finding a lattice of formal concepts. One problem of this method is that there is a huge number of formal concepts and most of them do not correspond to real concepts from the problem domain. Another problem is still not clear how to reason about data in this formal setting.

## 3. An Illustration – Inference in 2-Dimensional Space

### 3.1. Data in 2-dimensional space

Let $Z$ be a 2-dimensional space defined as the Cartesian product $Z = X \times Y$ of two sets $X$ and $Y$. It is a set of points $z = \langle x, y \rangle \in Z$ characterized by two coordinates taken along their axes: $x \in X$ and $y \in Y$.

In multidimensional data modeling, points in $Z$ and coordinates in $X$ and $Y$ are considered data elements. For example, a database in Fig. 3 consists of 10 data elements in three sets: $X = \{x_1, x_2, x_3\}$, $Y = \{y_1, y_2, y_3\}$ and $Z = \{z_1, z_2, z_3, z_4\}$. Elements in $X$ could describe existing colors and elements in $Y$ could describe possible prices while the products, characterized by colors and prices, are stored in $Z$. We can add elements, delete elements or update elements in this database and the only constraint is imposed by the structure of the multidimensional space. In our example it is a 2-dimensional space where data elements from $Z$ are tuples with two constituents: the first is taken from $X$ and the second is taken from $Y$. This structural constraint must be obeyed and there is no other way to define a data element in $Z$ except for choosing two existing coordinates along $X$ and $Y$. (We also assume that some coordinates can be optional which is denoted by a special marker like NULL instead of a concrete value).

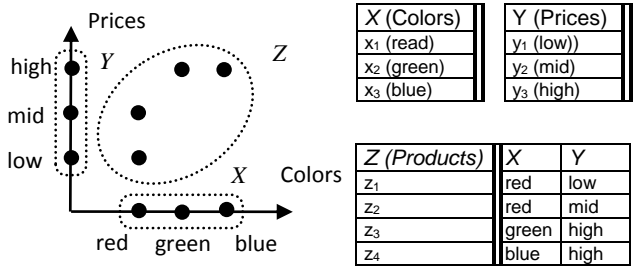

*Figure 3. An example of 2-dimensional database*

In the relational model, this structure can be expressed via foreign key constraints. In object data models, it is imposed via referential integrity. In star schema, space *Z* is interpreted as a master table while *X* and *Y* are interpreted as slave or detail tables. In OLAP, *Z* is a fact table while *X* and *Y* are dimension tables. In the concept-oriented model, *Z* is a lesser collection while *X* and *Y* are greater collections within a partially ordered set.

### 3.2. Inference in 2-dimensional space

Multidimensional view on data provides many advantages most of which have found their applications in OLAP, data warehousing and analytical applications. Yet, one property which has not been exploited before and is the main topic of this paper is the possibility to use multidimensional structure for inference. By inference we mean a possibility to derive constraints in one part of the model given constraints in another part of the model. Assume that we select some points along *X*. The main question is how to automatically find related points in *Z* and along *Y*?

The core idea of inference in multidimensional space is based on the assumption that points establish a relationship or dependency between their coordinates. If an element has some coordinates then they are considered somehow related. This relationship between coordinates is then used for inference. In the above example (Fig. 3), $z_1 = \langle x_1, y_1 \rangle$ which means that $z_1$ represents a relationship or dependency between $x_1$ and $y_1$. As a result, 'red' color and 'low' price are not independent elements but rather are connected via point $z_1$. A set of points in *Z* establishes a relationship between *X* and *Y* which in the presence of *Z* are not independent dimensions anymore.

Once points represent dependencies, they can be used to infer related coordinates along one axis given coordinates along another axis. For example, if we select 'red' color and then ask the question about possible prices (related price elements) then obviously it is 'low' or 'mid' (not 'high'). To draw this conclusion we have taken into account that there are only two points <'red', 'low'> and <'red', 'mid'> which relate the 'red' element with other coordinates in *Y*. We also see that green and blue products seem to be more expensive that red products. Our task is to formalize these observations and develop means for employing such kind of dependencies in various data analysis tasks.

This approach is analogous to having a numeric dependency between two axes represented as a function $y = z(x)$. Here function *z* represents a set of points in 2-dimensional space as some curve but the idea is the same: if we select some points $X' \subseteq X$ along *x* then we can infer related points $Y' \subseteq Y$ along *y* in two steps (Fig. 4): 1) select $Z' \subseteq Z$ where $Z' = \{\langle x, y \rangle \mid x \in X', y = z(x)\}$ is a subset of points on the plane, and 2) select $Y' \subseteq Y$ where $Y' = \{y \mid y \in Y, \langle x, y \rangle \in Z'\}$.

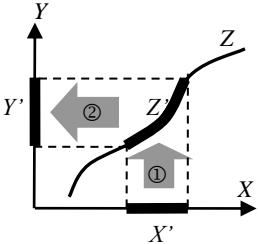

*Figure 4. Inference in 2-dimensional space using a function*

Importantly, this procedure relies on only the structure of dimensions with no use of inference rules. The role of inference rules is played by the same data in the database, that is, one data is used as dependencies to derive other data. In the next section we introduce a formal setting which allows us to represent multidimensional hierarchical spaces where coordinates are also treated as points with their own coordinates.

## 4. Concept-Oriented Model

### 4.1. Tuples and Concepts

In most existing data models, data elements are supposed to be formally represented by *tuples* denoted by angle brackets $\langle \ldots \rangle$ enclosing their members. A tuple consisting of *n* members is called *n*-tuple and *n* is called its arity or dimensionality. Tuple members are either primitive (atomic) values or other tuples. Each tuple member has a unique position which is referred to as a (simple) *dimension*. If *x* is a dimension then tuple where it is defined is called a *source*, $\mathrm{Src}(x)$, and the member represented by this dimension is called a *destination*, $\mathrm{Dst}(x)$. If $e = \langle x:a, y:b, z:c, \ldots \rangle$ is a source tuple then *x*, *y* and *z* are dimensions and $a = \mathrm{Dst}(x)$, $b = \mathrm{Dst}(y)$ and $c = \mathrm{Dst}(z)$ are their destinations. A *complex dimension* is a sequence of simple dimensions $x_1.\cdots.x_k$ where each next dimension starts where the previous dimension ends, $\mathrm{Src}(x_i) = \mathrm{Dst}(x_{i-1})$, $i = 2,3,\ldots,k$. The number *k* of constituents in a complex dimension is referred to as a *rank*.

Representing data elements by tuples has one drawback: it is not clear *how* tuple membership is implemented, that is, we do not know how one tuple is made a member of another tuple. There exist two ways for implementing tuple membership:

- [by-value] members are completely included as a whole without a possibility to share them among many tuples

- [by-reference] members are included using a unique reference (tuple-id) so that one member tuple can be shared among many parent tuples

Although both mechanisms are very important in data modeling, tuples as a formal representation method do not provide any indication which of them is used. Yet, even if one of the two mechanisms is explicitly specified, the problem is that in many cases we would like to freely vary between them. In other words, we would like to specify which dimensions of this tuple will be included by-value and which will be included by-reference.

A principled solution to this problem is provided by COM in its *duality principle* which defines an element as a couple of two tuples: one *identity tuple* and one *entity tuple*. Identity tuples are always passed by-value and serve as representatives of their entity counterparts. Entity tuples are always represented by-reference where the reference is their identity tuple. This separation is one of the corner stones of concept-orientation used in both data modeling and programming.

To model element types COM introduces a new construct, called *concept*, which is defined as a couple of two classes: one identity class and one entity class. Both classes may have arbitrary structure and their fields are referred to as dimensions. For example, books could be described by the following concept:

```
CONCEPT Books
  IDENTITY
    CHAR(10) isbn
  ENTITY
    CHAR(256) title
```

Importantly, identity class is not analogous to primary keys because a primary key consists of entity attributes manipulated like all other attributes. Identities can be thought of as user-defined surrogates or object references. If identity class is empty then concepts are equivalent to conventional classes. If entity class is empty then concepts describe value types. In the general case, concepts allow us to freely vary how things are represented and what data is passed by-value and by-reference. In particular, it generalizes the object-relational model (ORM) by modeling both value types (domains) and entity (relation) types using one construct without distinguishing between value-typed and relation-typed attributes.

### 4.2. Tuples for Representing Partial Order

A unique distinguishing feature of COM is that a set of elements is assumed to be a partially ordered set (poset). What is also new is that this partial order is represented by tuples themselves. Specifically, tuple membership is assumed to induce a partial order relation '<' (less than) which is a binary relation satisfying the conditions of irreflexivity and transitivity. This assumption is formulated as a tuple ordering principle:

[tuple ordering] $\langle \ldots, e, \ldots \rangle <_1 e$

Here $<_1$ means 'immediately less than' relation ('less than' of rank 1) and if $a < b$ then $a$ is referred to as a *lesser element* and $b$ is referred to as a *greater element*. Thus tuple members are supposed to be immediately greater than the tuples they are included in. And conversely, a tuple is immediately less than any its member tuple it is composed of. In terms of the cover relation, an element is said to cover all tuples where it is a member.

Since tuple membership is implemented via references (which are identity tuples), this principle essentially means that an element references its greater elements. In contrast, most other models use references only for connectivity within an arbitrary graph. References, and hence the induced partial order, have the following semantic interpretations in COM:

- [containment] (i) an element is a set of its lesser elements which reference it, and (ii) an element references sets (greater elements) where it is a member

- [specific-general] (i) an element is more specific than its greater (referenced) elements, and (ii) an element is more general than elements that reference it (lesser elements)

- [attribute-value] (i) an element is a value taken by an attribute of its lesser element, and (ii) an element is an object characterized by its greater elements as attribute values

- [multidimensional] (i) an element is a coordinate for its lesser element (interpreted as points), and (ii) an element is a point for its greater elements (interpreted as its coordinates)

These interpretations allow us to use COM as a unified model. However, in this paper we focus on the last interpretation only where partial order is treated as a multidimensional space which is described in the next section.

It is frequently convenient to consider a lattice of elements which (in finite case) has two special elements. The greatest element $g$, called *top*, is greater than any other element of the set: $\forall a \in R$, $a < g$. The least element *l*, called *bottom*, is less than any other element of the set: $\forall a \in R$, $l < a$. The number of immediate greater elements of this element is referred to as *arity* (also dimensionality, intention or valency). The number of immediate lesser elements of this element is referred to as *cardinality* (also extension). Fig. 5 is an example of a poset graphically represented using a Hasse diagram where an element is drawn under its immediate greater elements and is connected with them by edges.

At the level of concepts, tuple order principle means that dimension types specify greater concepts. Then a set of concepts is a poset where each concept has a number of greater concepts represented by its dimension types and a number of lesser concepts which use this concept in its dimensions. For example, assume that each book has one publisher:

```
CONCEPT Books // Books < Publishers
  IDENTITY
    CHAR(10) isbn
  ENTITY
    CHAR(256) title
    Publishers publisher // Greater concept
```

According to the order principle, `Publishers` is a greater concept.

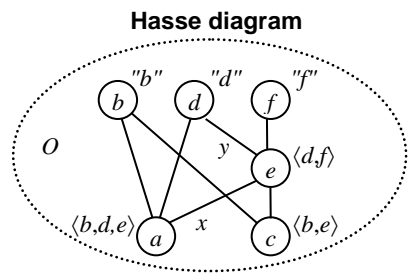

*Figure 5. Database is a partially ordered set*

### 4.3. Representing Hierarchical Multidimensional Space

In the context of this paper, the most important property of partial order is that it can be used for representing multidimensional hierarchical spaces (Savinov, 2005a). The basic idea is that greater elements are interpreted as coordinates with respect to their lesser elements which are interpreted as points. Multidimensionality appears due to the use of tuples each having multiple constituents. The hierarchical character is a consequence of treating coordinates (greater elements) as normal points with their own coordinates. In this way any point can be expanded into a set of primitive values.

Having only posets is not enough for representing multidimensional spaces. The reason is that we do not have the notion of axis and all elements exist in one large space like in (Raymond, 1996). In multidimensional space, any coordinate belongs to some axis and any point belongs to some set. The question is how the notions of axes and sets can be formally represented within the order-theoretic setting. To solve this problem we assume that a poset consists of a number of subsets, called *domains*: $O = X_1 \cup X_2 \cup \ldots \cup X_m$, $X_i \cap X_j = 0$, $\forall i \neq j$. Domains are interpreted as sets of coordinates (axes) or sets of points (spaces) and any element is included in some domain: $e \in X_k \subset O$.

Importantly, domains are also partially ordered and this structure is represented by tuples so that a domain is defined as a tuple consisting of its immediate greater domains: $X = \langle X_1, X_2, \ldots, X_n \rangle$, $X <_1 X_i$. Thus any element participates in two structures simultaneously: (i) it is included in some domain via the membership relation '$\in$', and (ii) it has some greater and lesser elements via the partial order relation '<'. (In fact, the first structure generalizes inheritance which is not considered in this paper but plays an important role for other mechanisms.) The question is how these two structures are related. In classical multidimensional space $M = X_1 \times X_2 \times \ldots \times X_n$, any point $p = \langle x_1, x_2, \ldots, x_n \rangle \in M$ may take its coordinates only from the corresponding domain: $x_i \in X_i$, $i = 1, 2, \ldots, n$. In the case of nested partially ordered sets this rule is called *type constraint* and is formulated as follows:

[type constraint] $e \in D \subset O \Rightarrow e.x \in D.x$

Here *e.x* is a greater element of *e* along dimension *x* and *D.x* is a greater domain of *D* along the same dimension *x*. This constraint means that an element may take its greater elements only from the greater domains. Such a set where each element participates in two structures simultaneously can be represented as a nested Hasse diagram. For example, neseted poset set *O* in Fig. 6 represents a multidimensional space shown in Fig. 3. This set consists of 3 domains *X*, *Y* and *Z* where $Z = \langle X, Y \rangle$ which means that *Z* has two greater domains *X* and *Y*. According to the type constraint, elements from *Z* may take their greater elements only in *X* and *Y*. For example, element $z_1$ is defined as a tuple $\langle x_1, y_1 \rangle$, that is, $x_1$ and $y_1$ are greater elements for $z_1$.

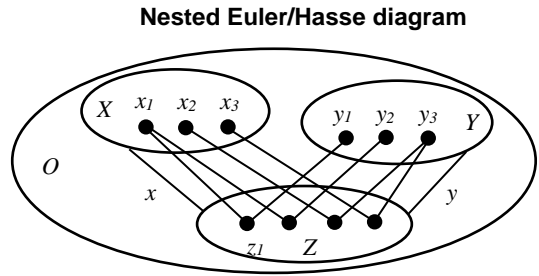

*Figure 6. An example of a nested partially ordered set*

A *schema* is defined as a partially ordered set of concepts where greater concepts are types of this concept dimensions. A *database schema* is defined as a partially ordered set of domains (collections) where elements within each domain have a type of some concept from the schema. A *database* is a partially ordered set of elements each belonging to one collection and having a number of greater elements referenced from its dimensions.

## 5. Inference in Hierarchical Multidimensional Space

### 5.1. Projection and De-Projection

Geometrically, projection of a set of points is a set of their coordinates. De-projection is the opposite operation where we select some coordinates and then find all points that have them. For example, points *d*, *e* and *f* in Fig. 7 are projected to coordinates *a* and *b*, and vice versa, coordinates *a* and *b* are de-projected to points *d*, *e* and *f*.

In terms of partial order, projection means finding all greater elements and de-projection means finding all lesser elements for the selected set of elements. Fig. 8 illustrates projection and de-projection for the example shown in Fig. 7 but representing it in terms of partial order relation where projection means moving up and de-projection means moving down. Taking into account that greater elements are represented by references, projection is a set of elements referenced by the selected elements and de-projection is a set of elements which reference the selected elements.

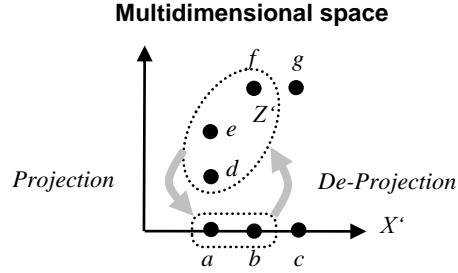

*Figure 7. Projection and de-projection in multidimensional space*

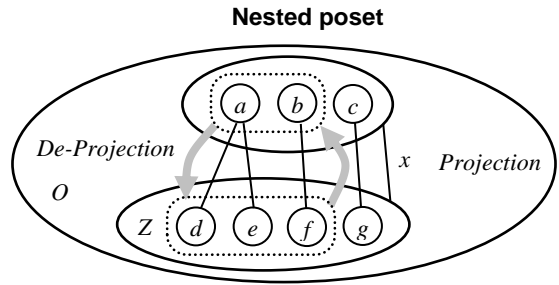

*Figure 8. Projection and de-projection in a poset*

To formally describe these two operation we need to select a subset $Z'$ of points from the source domain, $Z' \subset Z$, and then choose some (complex) dimension $d$ with the destination in the target greater domain, $\mathrm{Dst}(x) = D > Z$. Then projection, denoted as $Z' \to x \to D$, returns a subset of elements $D' \subset D$ which are greater than elements from $Z'$ along dimension $d$:

[projection] $Z' \to x \to D = D' = \{d \in D \mid z <_x d, z \in Z'\} \subset D$

Note that any element can be included in projection only one time even if it has many lesser elements. In terms of tuples, this definition means that projection returns all tuple members along the specified dimension:

$Z' \to x \to D = D' = \{d \in D \mid z = \langle \ldots, x : d, \ldots \rangle \in Z'\} \subset D$

De-projection is the opposite operation which returns all points from the target domain $F$ which are less than the elements from the selected source subset $Z'$:

[de-projection] $Z' \leftarrow x \leftarrow F = F' = \{f \in F \mid f <_x z, z \in Z'\} \subset F$

In terms of tuples, de-projection $Z' \leftarrow x \leftarrow F$ returns all tuples which include the selected tuples as members:

$Z' \leftarrow x \leftarrow F = F' = \{f \in F \mid f = \langle \ldots, x : z, \ldots \rangle, z \in Z'\} \subset F$

In the concept-oriented query language (COQL) (Savinov, 2014a, 2011b), a set of elements is written in parentheses with constraints separated by bar symbol. For example, `(Books | price < 10)` is a set of all cheap books. (So parentheses are like SQL FROM and bar symbol is like SQL WHERE.) Projection operation is denoted by right arrow `'->'` followed by a dimension name which is followed by the target collection (Savinov, 2012d). In the database schema, projection means moving up to the domain of the specified dimension (Fig. 9). It returns all (greater) elements which are referenced by the selected elements.

For example (Fig. 10), all publishers of cheap books can be found by projecting them up (right arrow) to the `Publishers` collection along the `publisher` dimension:

```
(Books | price < 10) -> publisher -> (Publishers)
```

De-projection is the opposite operation denoted by left arrow `'<-'`. It returns all (lesser) elements which reference elements from the selected source collection. For example, all books published by a selected publisher can be found as follows (Fig. 10):

```
(Publishers | name=="XYZ") <- purchase <- (Books)
```

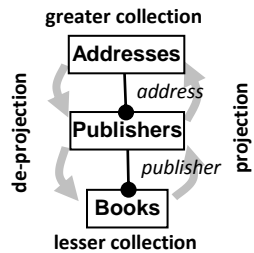

*Figure 9. Projection and de-projection in database schema*

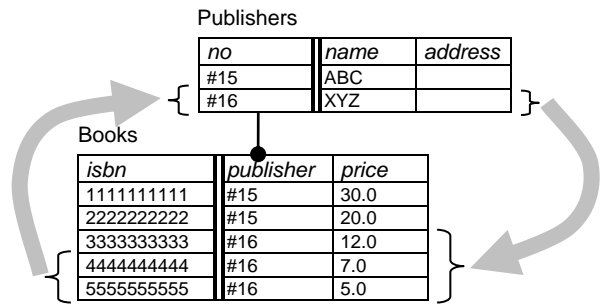

*Figure 10. An example of projection and de-projection*

Projection and de-projection operations have two direct positive properties: they eliminate the need in join and group-by operations. Joins are not needed because sets (domains) are connected using multidimensional hierarchical structure of the model (Savinov, 2005b). Group-by is not needed because an element is interpreted as a group consisting of its lesser elements which can be accessed using the structure. Given an element (group) we can get its members by applying de-projection operation (Fig. 11). For example, if it is necessary to select only publishers with more than 10 books published then it can be done it as follows:

```
(Publishers | COUNT(publisher <- (Books)) > 10)
```

Here de-projection `publisher <- (Books)` returns all books of this publisher and then their count is compared with 10. This query can be modified by counting only expensive books:

```
(Publishers | SUM(publisher <- (Books | price > 100)) > 10)
```

Specifying queries in business terms is simpler and more natural than using explicit joins and group-by especially for complex analytical queries which involve many collections from different parts of the model, complex constraints and nested groupings.

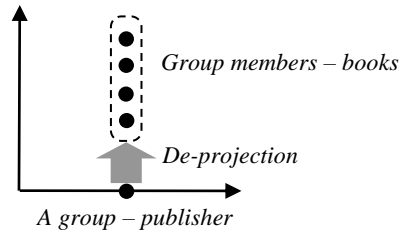

*Figure 11. Group is a coordinate and its members are points*

### 5.2. Constraint Propagation

A query can be defined as consisting of constraints and propagation rules. Constraints specify conditions for selecting elements from the domain. Propagation rules indicate how constraints imposed on one domain are used to select elements from another domain. For example, we can select a book title, publishing date or price while the task is to find all publishers which satisfy these selected values. However, the source constraints and the target elements belong to different parts of the model and propagation rules allow us to connect them.

There are two opposite ways how a propagation path can be specified:

- [From target to source] It is a traditional approach where we specify how target elements depend on the primitive elements. We start from the target element structure and then reduce the description to the source domains where constraints are imposed. For example (Fig. 9), if we need to select all books published in Germany then it is done as follows: `(Books | publisher.address.country = 'DE')`. Note that we start from the book (target) and then specify dimensions leading to the primitive domain (country code). For each book element, we can determine whether it satisfies the constraints or not.

- [From source to target] Here we start from the source constraints and specify the path to the target domain along with constraint propagation rules and operators. For example (Fig. 9), given a country code we can find all books as follows: `'DE' <- country <- address <- publisher <- (Books)`. Here we start from the primitive domain (country codes) and then proceed to the target (books).

The main benefit of having projection and de-projection operations is that they allow us to easily propagate arbitrary constraints through the model by moving up (projection) and down (de-projection). For example, given a collection of `Books` we can find all related addresses by projecting it up to the `Addresses` collection (Fig. 9):

```
(Books | price < 10)
  -> publisher -> (Publishers)
  -> address -> (Addresses)
```

In most cases either intermediate dimensions or collections can be omitted so that the query remains unambiguous but looks even simpler:

```
(Books | price < 10) -> publisher -> address
```

```
(Books | price < 10) -> (Publishers) -> (Addresses)
```

De-projection allows us to move down through the partially ordered structure of the model which is interpreted as finding group members. For example, given a country code we can find all related `Books` using the following query (Fig. 9):

```
(Addresses | country == 'DE')
  <- address <- (Publishers)
  <- publisher <- (Books)
(Addresses | country == 'DE')
  <- (Publishers) <- (Books)
```

Note that such queries are very useful for nested grouping which are rather difficult to describe using the conventional group-by operator.

Constraint propagation can be further simplified if instead of a concrete dimension path we specify only source and target collections. The system then reconstructs the propagation path itself. Such projection and de-projection with an undefined dimension path will be denoted by `'*->'` and `'<-*'` (with star symbol interpreted as *any* dimension path). The previous queries can be then rewritten as follows:

```
(Books | price < 10) *-> (Addresses)
```
```
(Addresses | country == 'DE') <-* (Books)
```

Such queries are especially useful when the model structure is either too complicated or changes frequently. The main restriction is that constraints can be automatically propagated only in one direction (up or down). In the next section we eliminate this limitation so that constraints can be propagated automatically to any part of the model.

## 5.3. Inference

Automatically propagating constraints only up or down is a restricted version of inference because only more general (projection) or more specific (de-projection) data can be derived. If source and target collections have arbitrary positions in the schema then this approach does not work because they do not belong to one dimension path that can be used for projecting or de-projecting. In the general case, constraint propagation path consists of more than one projection and de-projection steps leading from the source collection to the target. Of course, this propagation path can be specified explicitly as part of the query but our goal is to develop an *automatic* procedure for finding related items in the database which is called inference.

The main idea of the proposed solution is that source and target collections have some common lesser collection which is treated as a relationship or *dependency* between them as described in Section 3.2. In this case, constraints imposed on the source collection can be used to select a subset of elements from this lesser collection using de-projection. And then the selected elements are used to constrain the target elements using projection. In OLAP terms, we impose constraints on dimension tables, propagate them down to the fact table, and then finally use these facts to select values from the target dimension table. In terms of multidimensional space this procedure means that we select some points along one axis, then de-project them to the plane by selecting a subset of points, and finally project these points to the target axis and use these coordinates as the result of inference.

If *X* and *Y* are two greater collections, and *Z* is their common lesser collection then the proposed inference procedure consists of two steps:

1. [De-projection] Source constraints $X' \subset X$ are propagated down to the set $Z$ using de-projection: $Z' = X' \leftarrow * \leftarrow Z \subset Z$

2. [Projection] The constrained set $Z' \subset Z$ is propagated up to the target set $Y$ using projection: $Y' = Z' \rightarrow * \rightarrow Y \subset Y$

Here by star symbol we denote an arbitrary dimension path. In the case of $n$ independent source constraints $X_1', X_2', \ldots, X_n'$ imposed on sets $X_1, X_2, \ldots, X_n$ the de-projection step is computed as an intersection of individual de-projections: $Z' = \bigcap X_i' \leftarrow * \leftarrow Z$.

In COQL, inference operator is denoted as `<-*->` (de-projection step followed by projection step via an arbitrary dimension path). It connects two collections from the database and finds elements of the second collection which are related to the first one. To infer the result, the system chooses their common lesser collection and then builds de-projection and projection dimensions paths. After that, inference is performed by propagating source constraints to the target along this path. For example (Fig. 12), given a set of young writers we can easily find related countries by using only one operator:

```
(Writers | age < 30) <-*-> (Addresses) -> countries
```

To answer this query, the system first chooses a common lesser collection, which is `WriterBooks` in this case, and then transforms this query to two operations of de-projection and projection:

```
(Writers | age < 30)
  <-* (WriterBooks) // De-projection
  *-> (Addresses) -> countries // Projection
```

After that, the system reconstructs complete constraint propagation paths:

```
(Writers | age < 30)
  <- writer <- (WriterBooks)
  -> book -> (Books)
  -> publisher -> (Publishers)
  -> address -> (Addresses) -> countries
```

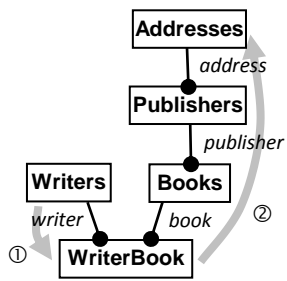

*Figure 12. Inference via common lesser collection*

In the case of many dependencies (common lesser collections) or many de-projection/projection paths between them, the system propagates constraints using all of them. This means that all source constraints are first propagated down along all paths to all lesser collections using de-projection. After that, all the results are propagated up to the target collection using all existing dimension paths.

If the user wants to customize inference and use only specific dimensions or collections then they can be provided as part of the query. For example, assume that both `Publishers` and Writers have addresses (Fig. 13). Accordingly, there are two alternative paths from source to target and two alternative interpretations of the relationship: writers living in some country and writers publishing in this country. This ambiguity can be explicitly resolved in the query by specifying the required common collection to be used for inference:

```
(Addresses | country == 'DE') <-* (WriterBooks) *-> (Writers)
```

In this way, we can solve the problem of having multiple propagation paths. In the next section we consider the problem of having no propagation path between source and target collections.

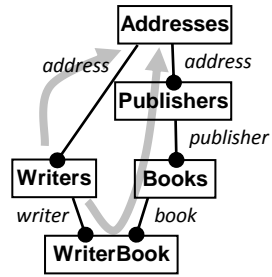

*Figure 13. Two alternative inference paths*

### 5.4. Use of Background Knowledge

If the model has a bottom collection which is less than any other collection then it can be used as a dependency between *any* pair of source and target collections and hence inference is always possible. Yet, if the model does not have a bottom collection then the question is how to carry out inference in this case. Formally, collections which do not have a common lesser collection are independent, that is, their elements are unrelated. Since source constraints cannot reach the target, we will always get all elements from the target collection.

For example (Fig. 14), if books are being sold in different shops then the model has two bottom collections: WriterBooks and Sellers. Assume now that it is necessary to find all shops related to a set of writers:

```
(Writers | age < 30) <-*-> (Shops) // All shops
```

The propagation path should go through a common lesser collection which is absent in this example and therefore inference is not possible.

One solution to this problem is to formally introduce a bottom collection which is equal to the Cartesian product of its immediate greater collections. In COQL, this operation is written as a sequence of collections in parentheses separated by comma:

```
Bottom = (WriterBooks, Sellers)
```

However, this artificial bottom collection (shown as a dashed rectangle in Fig. 14) does not impose any constraints and hence Writers and Shops are still independent.

To get meaningful results we have to impose additional constraints on the bottom collection. These constraints represent implicit dependencies between data elements, called background knowledge. They can be expressed as any condition which selects a subset of elements from the bottom collection. In particular, it can be represented as some dependency between its attributes. In our example, we assume that a written book is the same as a sold book:

```
Bottom = (WriterBooks wb, Sellers s | wb.book == s.book)
```

Now the `Bottom` collection contains only a subset of the Cartesian product of its two greater collections and can be used for inference. We simply specify this bottom collection as part of the query:

```
(Writers | age < 30)
  <-*                                                    // De-project
  (WriterBooks bw, Sellers s| bw.book == s.book)
  *-> (Shops)                                            // Project
```

Here the selected writers are de-projected down to the bottom collection. Then this constrained bottom collection is propagated up to the target. As a result, we will get all shops with books written by the selected authors. Note how simple this query is (especially in comparison with its SQL counterpart which has to contains many joins and explicit intermediate tables). What is more important, it is very natural because we specify what we can get rather than how the result set has to be built.

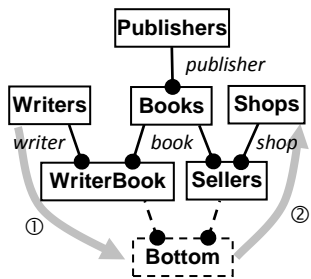

*Figure 14. Use of background knowledge for inference*

## 6. Conclusion

In this paper we have described the idea of having inference capabilities as an inherent part of multidimensional data models and analytical query languages. It is a highly non-trivial task because multidimensional modeling has been traditionally aimed at numeric analysis rather than reasoning. The proposed solution is based on the observation that constraints imposed on one axis can be propagated to another axis using points on the plane as dependencies. Another observation is that hierarchical multidimensional spaces can be represented by means of the partial order relation. To formalize these observations, we introduced the notion of nested partially ordered set with operations of projection and de-projection. These two operations are then used to propagate constraints through the hierarchical and multidimensional structure of the model. The proposed inference procedure is based on the assumption that lesser elements describe dependencies between greater elements. In this case, inference is defined as a two-step procedure where source constraints are propagated down via de-projection and then they are propagated up to the target via projection.

This work is a step towards developing a unified model which provides equal support for transactional, analytical and reasoning operations. The main benefit of the proposed approach is that now inference can be made integral part of multidimensional databases and analytical applications. If a standard OLAP application allows the user to find out how some numeric measure is distributed over several dimensions then now it is also possible to infer data items along one dimension by constraining items along other dimensions. This method can be also implemented at the level of query languages and then inference will be one of the data manipulation operators which can be combined with other operators for carrying out complex ad-hoc analysis tasks.